\documentclass[onecolumn,showpacs,fleqn,nobibnotes]{revtex4}

\usepackage{amsmath}
\usepackage{graphicx}
\usepackage{float}
\usepackage{subfigure}

\def\lsim{\raise0.3ex\hbox{$<$\kern-0.75em\raise-1.1ex\hbox{$\sim$}}}
\def\gsim{\raise0.3ex\hbox{$>$\kern-0.75em\raise-1.1ex\hbox{$\sim$}}}
\def\pom{{I\!\!P}}
\newcommand{\beq}{\begin{equation}}
\newcommand{\eeq}{\end{equation}}

\begin{document}

\title{Meson production in two-photon interactions at energies available at CERN Large Hadron Collider}
\author{V.P. Gon\c{c}alves, D. T. da Silva and W. K. Sauter}

\affiliation{Instituto de F\'{\i}sica e Matem\'atica, Universidade Federal de
Pelotas\\
Caixa Postal 354, CEP 96010-900, Pelotas, RS, Brazil.}

\begin{abstract}
The meson production cross sections  are estimated considering photon-photon interactions in hadron - hadron collisions at  CERN LHC energies. We consider a large number of mesons with photon-photon partial decay width
 well constrained by the experiment and some mesons which  
 are currently considered as  hadronic molecule and glueball candidates. Our results demonstrate  that the experimental analysis of these states is feasible at CERN - LHC.

\end{abstract}
\maketitle

The Large Hadron Collider (LHC) at CERN  started high energy collisions two years ago. During this 
period a large amount of data have been collected considering $pp$ collisions at $\sqrt{s}$ = 0.9, 2.36 and 7 TeV as well as $PbPb$ collisions at $\sqrt{s}$ = 2.76 TeV.  Currently, there is a great expectation that LHC will discover  new physics beyond the Standard Model, such as 
supersymmetry or extra dimensions. However, we should remember that 
the LHC opens a new kinematical regime at high energy, where several questions related to the 
description of the Quantum
Chromodynamics (QCD) remain without  satisfactory answers. 
Some open questions  are the search for non-$q\bar{q}$
resonances, the determination of the spectrum of  $q\bar{q}$ states and the 
identification of states with anomalous $\gamma \gamma$ couplings. A possible 
way to study these problems is the study of meson production in two-photon interactions \cite{brodsky,achasov}. 
In general, this process is studied in leptonic colliders. 
An alternative is to use ultra-relativistic protons and nuclei, which  give rise 
to strong electromagnetic fields and estimate the production of a given 
final state considering the photon - photon and photon - hadron 
interactions. In particular, it is possible to study photon - photon 
interactions in proton - proton and nucleus - nucleus collisions at LHC (For a review see Ref. \cite{baur}). 
Recently, Bertulani \cite{bertulani}  revisited this subject and  proposed the study of the meson production in ultraperipheral heavy ion collisions at LHC in order to constrain the two-photon decay widths. In this letter we extend this previous study for the meson production in two-photon interactions in proton - proton collisions. Initially we calculate the cross sections for mesons
with photon-photon partial decay width
 well constrained by the experiment, which allows to constrain the theoretical methods and calibrates the experimental techniques. After we predict the cross sections for mesons 
which are currently considered as glueball and hadronic molecule candidates. Our results shows that LHC can be used to investigate  these states.

Lets consider the hadron-hadron interaction at large impact parameter ($b > R_{h_1} + R_{h_2}$) 
and at ultra relativistic energies (For recent reviews see, e.g., Ref. \cite{uphic}). In this regime we expect the electromagnetic interaction to 
be dominant.
In  heavy ion colliders, the heavy nuclei give rise to strong electromagnetic fields due to the 
coherent action of all protons in the nucleus, which can interact with each other. In a similar 
way, it also occurs when considering ultra relativistic  protons in $pp(\bar{p})$ colliders.
The photon stemming from the electromagnetic field
of one of the two colliding hadrons can interact with one photon of
the other hadron (two-photon process) or can interact directly with the other hadron (photon-hadron
process). The total cross section for a given process can be factorized in terms of 
the equivalent flux of photons of the hadron projectile and  the photon-photon or photon-target 
production cross section. In the case of the production  of a neutral state $X$ in two-photon 
interactions the total cross section is given by (See, e.g., \cite{serbo})
\begin{eqnarray}
\sigma(h_1 h_2 \rightarrow h_1 \otimes X \otimes h_2)=\int dx_{1} \int dx_{2}
f_{h_1}^{\gamma}(x_{1}) f_{h_2}^{\gamma}(x_{2})\sigma_{\gamma\gamma}^{X}(x_{1}x_{2}s)
\label{totalcs}
\end{eqnarray}
where $\otimes$ characterizes the presence of a rapidity gap in the final state, $s$ is the 
squared center of mass energy, $f_{h_i}^{\gamma}$ is the distribution function which is 
associated to the flux of photons generated by the hadron $h_i$ ($i = 1,2$), $x_i = \omega_i/E_i$, 
with $\omega_i$ and $E_i$ the photon and hadron energies, respectively.
Moreover, $\sigma_{\gamma\gamma}^{X}$  is the photon-photon cross section given by
\begin{eqnarray}
\sigma_{\gamma\gamma}^{X}(x_{1}x_{2}s)=8\pi^{2}(2J+1)\frac{\Gamma_{X
\rightarrow\gamma\gamma}}{m_{X}}\delta(x_{1}x_{2}s-m_{X}^{2}) \,\,,
\label{sigma-foton}
\end{eqnarray}
where $J$, $m_X$ and $\Gamma_{X
\rightarrow\gamma\gamma}$ are the spin, mass 
and the photon-photon partial decay width of the final state $X$, respectively, and the $\delta$ function enforces energy conservation.

The main input in our calculations are the equivalent photon flux for a ultrarelativistic proton, $f^{\gamma}(x)$,  and the two photon partial decay widths, $\Gamma_{X
\rightarrow\gamma\gamma}$. Currently there are different models for the equivalent photon flux available in the 
literature (See e.g. Ref. \cite{nystrand}). The general expression for the equivalent photon flux of an extended object is given by \cite{serbo}
\begin{eqnarray}
f^{\gamma}(x) = \frac{\alpha Z^2}{\pi} \frac{1 - x + 0.5x^2}{x} \int_{Q^2_{min}}^{\infty} dQ^2 \frac{Q^2 - Q^2_{min} }{Q^4} |F(Q^2)|^2 \,\,,
\label{fluxgeral}
\end{eqnarray}
where $Q^2$ is the momentum transfer from the projectile and $F(Q^2)$ its form factor. Moreover, $Q^2_{min} \approx (x M_A)^2/(1-x)$ with $M_A$ the mass of the projectile. 
The presence of the form factor cuts off the photon flux above $1 \simeq 2$ GeV$^2$. As its dependence on the photon virtuality  is $\approx 1/Q^2$, the average virtuality is very small, which allow us to treat the processes as due to quasi-real photon photon collisions.

\begin{figure}
\includegraphics[scale=0.2]{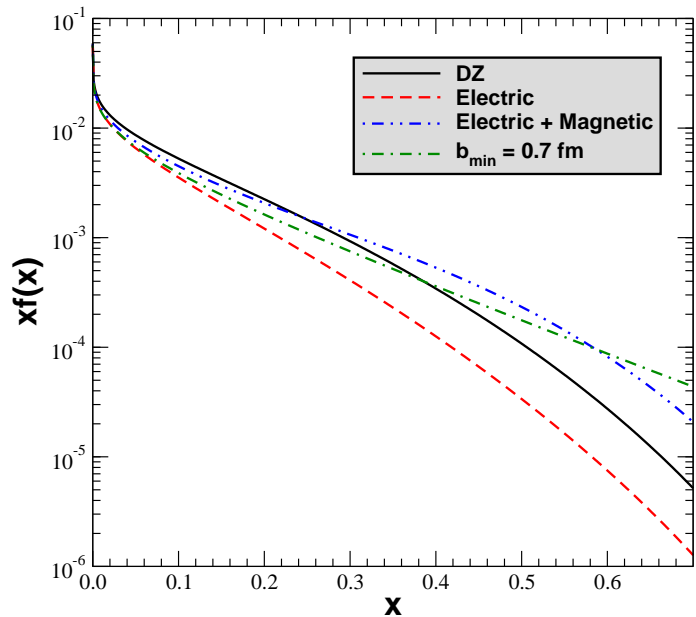}
\caption{Comparison between different models for the equivalent photon flux as a function of the fractional photon energy  $x$.}
\label{fig1}
\end{figure}

Considering only the electric dipole form factor for the proton, $F_E (Q^2) = 1/(1+Q^2/0.71\,$GeV$^2)^2$, the following expression for the equivalent photon flux can be obtained 
\begin{eqnarray}
f_h^{\gamma}(x)&=&\frac{\alpha}{\pi}\left(\frac{1-x+0.5x^{2}}{x}\right)
\left[\frac{A+3}{A-1}\ln(A)-\frac{17}{6}-\frac{4}{3A}+\frac{1}{6A^{2}}\right] \,\,,
\end{eqnarray}
where $A=1+(0.71 \mathrm{GeV}^{2})/Q_{min}^{2}$. We denote this model by {\it Electric} in what follows. 
If the term containing $Q^2_{min}$ in Eq. (\ref{fluxgeral}) is disregarded, the 
 equivalent photon spectrum of high energy protons
is given as follow
\begin{eqnarray}
f_h^{\gamma}(x)&=&\frac{\alpha}{\pi}\left(\frac{1-x+0.5x^{2}}{x}\right)\left[\ln(A)
-\frac{11}{6}+\frac{3}{A}-\frac{3}{2A^{2}}+\frac{1}{3A^{3}}\right]\,\,.
\end{eqnarray}
This expression was derived originally by Dress and Zeppenfeld in Ref. \cite{dz} and will be denoted DZ  hereafter.
In Ref. \cite{kniehl}, the author studied the effect of including the magnetic dipole moment and the corresponding magnetic form factor of the proton, obtaining a spectrum (denoted {\it Electric + Magnetic} hereafter) which is smaller than the DZ one at small $x$. Another model  considered  in literature is the use of a minimum impact parameter, $b_{min}$ = 0.7 fm, in  the photon energy spectrum produced by a point particle, which is given by
\begin{eqnarray}
f^{\gamma}(x) = \frac{\alpha Z^2}{\pi x} \left[ 2 \xi K_0 (\xi) K_1 (\xi) - \xi^2 ( K_1^2 (\xi) - K_0^2 (\xi) ) \right]
\end{eqnarray}  
where $K_0$ and $K_1$ are modified Bessel functions and $\xi \equiv x M_Ab_{min}$.
 In Fig. \ref{fig1}
we present a comparison between these different models for the photon flux. A basic characteristic of the distinct models for the  photon spectrum is that they diminish  with energy approximately like $1/x$.  Consequently, the photon spectrum is strongly peaked at low $x$, so that the photon - photon center of mass energy, $\approx 2 \sqrt{\omega_1 \omega_2}$, is much smaller than the center of mass energy of the proton - proton system. Therefore, the main contribution for the total cross section, Eq. (\ref{totalcs}) comes from the small $x$ behaviour of the photon spectrum. In this region, the DZ model predicts a larger photon flux. In contrast, the {\it Electric} one, predicts the lower photon flux. The other two models, the {\it Electric + Magnetic} and $b_{min}$ = 0.7 fm models, predict intermediate values for the photon flux. At low $x$ ($\le 0.05$) the difference between the models is ever smaller than 20 \%. However, the difference  increases at larger values of  $x$. 
 At $x = 0.1$ the difference among the DZ and {\it Electric} models is $\approx$ 25 \%, increasing for 
$\approx$ 100 \% at $x = 0.4$. 
In this letter we will use the $Electric$ and $DZ$ photon fluxes in our calculations, which allow us to estimate the theoretical uncertainty in our predictions.

\begin{table}[t]
\begin{center}
\begin{tabular}{|c|c|c|c|c|c|}
\hline 
State & $\Gamma_{\gamma\gamma}^{exp.}$(keV) & \multicolumn{2}{c|}{$\sigma_{Electric}$ (pb)} & \multicolumn{2}{c|}{$\sigma_{DZ}$ (pb)}\tabularnewline
\cline{3-6} 
 & $X\rightarrow\gamma\gamma$ & $7$ TeV & $14$ TeV & $7$ TeV & $14$ TeV\tabularnewline
\hline
\hline 
$\pi^{0}$ & $(8.3\pm0.49)\times10^{-3}$ & $2426.0$ & $3008.0$ & $2812.0$ & $3453.9$\tabularnewline
\hline 
$\eta$ & $0.510\pm0.026$ & $1368.9$ & $1760.0$ & $1624.8$ & $2062.9$\tabularnewline
\hline 
$\eta^{\prime}$ & $4.29\pm0.14$ & $1730.6$ & $2265.9$ & $2078.0$ & $2682.0$\tabularnewline
\hline 
$f_{0}(980)$ & $0.29_{-0.06}^{+0.07}$ & $108.0$ & $141.8$ & $130.0$ & $167.9$\tabularnewline
\hline 
$a_{0}(980)$ & $0.30\pm0.10$ & $111.9$ & $146.7$ & $134.0$ & $173.7$\tabularnewline
\hline 
$f_{2}(1270)$ & $3.03\pm0.35$ & $2300.8$ & $3043.9$ & $2781.0$ & $3623.7$\tabularnewline
\hline 
$a_{2}(1320)$ & $1.0\pm0.06$ & $677.8$ & $897.8$ & $820.0$ & $1069.6$\tabularnewline
\hline 
$f_{2}^{\prime}(1525)$ & $0.081\pm0.009$ & $33.0$ & $44.0$ & $40.0$ & $53.0$\tabularnewline
\hline 
$f_{2}(1565)$ & $0.70\pm0.14$ & $264.9$ & $353.0$ & $321.9$ & $422.0$\tabularnewline
\hline 
$a_{2}(1700)$ & $0.30\pm0.05$ & $79.6$ & $106.6$ & $97.0$ & $127.7$\tabularnewline
\hline 
$f_{2}(1750)$ & $0.13\pm0.04$ & $33.0$ & $44.0$ & $40.0$ & $52.9$\tabularnewline
\hline 
$\eta_{c}(1S)$ & $6.7_{-0.8}^{+0.9}$ & $58.0$ & $80.0$ & $72.0$ & $97.0$\tabularnewline
\hline 
$\chi_{c0}(1P)$ & $2.28\pm0.3$ & $11.0$ & $15.9$ & $14.0$ & $19.0$\tabularnewline
\hline 
$\chi_{c2}(1P)$ & $0.504\pm0.06$ & $11.0$ & $15.0$ & $13.7$ & $18.6$\tabularnewline
\hline 
$\eta_{c}(2S)$ & $1.30\pm0.6$ & $5.0$ & $7.0$ & $6.0$ & $8.9$\tabularnewline
\hline
\end{tabular}
\caption{Cross sections  for meson production at LHC energies considering the partial decay rates 
 given by the Particle Data Group \cite{PDG}.}
\label{I}
\end{center}
\end{table}

In what follows we present our predictions for the total cross section 
considering proton - proton collisions at LHC and center-of-mass energies of 
$7$ TeV and $14$ TeV. 
We consider $\Gamma_{\gamma \gamma}$ either taken from experiment or from theory.
Initially, we present in Table \ref{I} our predictions for the mesons which have partial 
decay ratio reasonably well constrained by the experiment, which allows to use the values 
present in the Particle Data Group \cite{PDG}. 
As emphasized before, the study of mesons which have its decay ratio well known 
is fundamental to constrain the theoretical methods and calibrate the experimental techniques, for in a second moment investigate the production of exotic particles.
As expected from Eqs. (\ref{totalcs}) and (\ref{sigma-foton}), the cross sections decrease at larger values of the meson mass $m_X$ and increase at larger values of $\Gamma_{\gamma \gamma}$. Moreover, the cross sections increase by  $\approx 30 \%$ when the center of mass energy increases from 7 to 14 TeV. The difference between the $Electric$ and $DZ$ depends of the final state and it is of the order of $ \approx 20 \%$ .
 Assuming the  design luminosity
${\cal L} = 10^7$ mb$^{-1}$s$^{-1}$ the corresponding event rates will be larger than $10^5$  events/year at $\sqrt{s} = 7$ TeV, making the experimental analysis of these final states feasible at LHC.

\begin{table}[t]
\begin{center}
\begin{tabular}{|c|c|c|c|c|c|}
\hline 
State & $\Gamma_{\gamma\gamma}^{th.}$(keV) & \multicolumn{2}{c|}{$\sigma_{Electric}$ (pb)} & \multicolumn{2}{c|}{$\sigma_{DZ}$ (pb)}\tabularnewline
\cline{3-6} 
 & $X\rightarrow\gamma\gamma$ & $7$ TeV & $14$ TeV & $7$ TeV & $14$ TeV\tabularnewline
\hline
\hline 
$\pi(1300)$ & $0.43$ & $61.10$ & $80.89$ & $73.90$ & $96.34$\tabularnewline
\hline 
$f_{4}(2050)$ & $0.36$ & $101.52$ & $136.80$ & $124.18$ & $164.47$\tabularnewline
\hline 
$\eta_{b}(1S)$ & $0.17$ & $0.024$ & $0.035$ & $0.031$ & $0.044$\tabularnewline
\hline 
$\chi_{b0}(1P)$ & $13.0\times10^{-3}$ & $0.0015$ & $0.0022$ & $0.0019$ & $0.0028$\tabularnewline
\hline 
$\chi_{b2}(1P)$ & $3.7\times10^{-3}$ & $0.0021$ & $0.0031$ & $0.0027$ & $0.0039$\tabularnewline
\hline
\end{tabular}
\caption{Cross sections and event rates for meson production considering theoretical decay rates presented in Ref.  \cite{bertulani}.  }
\label{II}
\end{center}
\end{table}

In Tables \ref{II}, \ref{III} and \ref{IV} we present our predictions for some mesons that does not have two photon partial decay rates well constrained by the experiment. 
Our goal now is to verify if the study of the meson production in two-photon interactions at LHC can be used to constrain the partial decay rates and, consequently, the theoretical models considered in our calculations. 
In Table \ref{II} we consider the theoretical values for $\Gamma_{\gamma \gamma}$ as given in \cite{bertulani}. The small values of $\Gamma_{\gamma \gamma}$ for the $\chi_{b0}$ and $\chi_{b2}$ states implies that the experimental analysis of these states is a hard task. 

Recently, several new observed states have been interpreted as being hadronic molecules, i.e bound states of two or more mesons (For a review see, e.g., \cite{review_mol}). In these models
two photon radiative decays are considered diagnostic tools which are sensitive to the inner structure of the short-lived molecules. Here we consider the model proposed in Refs. \cite{mol1,mol2,mol3}, where 
the decay rates are obtained considering an effective Lagrangian which includes both the coupling of the molecular bound state to their hadronic constituents and the coupling of the constituents to other hadrons and photons.
For instance, in this model the $X(3940)$ meson  is considered as a superposition of the molecular $D^{*+}D^{*-}$ and $D^{*0}D^{*0}$ states, while the $X(4140)$ meson is a bound state of $D_s^{*+}$ and $D_s^{*-}$ mesons.
In Table \ref{III} we present our predictions for some hadronic molecule candidates considering the two-photon decay rates given in Refs. \cite{mol1,mol2,mol3}.
We assume that the masses of 
the mesons  $f_{0}(1370)$, 
 $f_{0}(1710)$, $X(3940)$  
and $X(4140)$   
are given by   $1523$, 
 $1721$, $3943$ and $4143$ MeV, respectively.
For the lighter state, $f_{0}(1370)$,  we predict large values for the total cross sections and event rates larger than 
$10^7$  events/year at $\sqrt{s} = 7$ TeV. For the heavy $X$ states we predict event rates larger than 
$10^5$  events/year with a strong dependence on the quantum numbers of the state. The large values predicted by this model imply that  the experimental analysis of these final states can be useful to constrain the underlying physics.

\begin{table}[t]
\begin{center}
\begin{tabular}{|c|c|c|c|c|c|c|}
\hline 
State & Mass & $\Gamma_{\gamma\gamma}^{th.}$(keV) & \multicolumn{2}{c|}{$\sigma_{Electric}$ (pb)} & \multicolumn{2}{c|}{$\sigma_{DZ}$ (pb)}\tabularnewline
\cline{4-7} 
 & (MeV) & $H\rightarrow\gamma\gamma$ & $7$ TeV & $14$ TeV & $7$ TeV & $14$ TeV\tabularnewline
\hline
\hline 
$f_{0}(1370)$ & $1523$ & $1.3$ & $108.7$ & $144.1$ & $131.3$ & $172.2$\tabularnewline
\hline 
$f_{0}(1710)$ & $1721$ & $0.05$ & $2.7$ & $3.6$ & $3.3$ & $4.4$\tabularnewline
\hline 
$X(3940),\,0^{++}$ & $3943$ & $0.33\pm0.01$ & $1.01$ & $1.4$ & $1.3$ & $1.7$\tabularnewline
\hline 
$X(3940),\,2^{++}$ & $3943$ & $0.27\pm0.01$ & $4.1$ & $5.7$ & $5.1$ & $7.0$\tabularnewline
\hline 
$X(4140),\,0^{++}$ & $4143$ & $0.63\pm0.01$ & $1.6$ & $2.3$ & $2.02$ & $2.8$\tabularnewline
\hline 
$X(4140),\,2^{++}$ & $4143$ & $0.50\pm0.01$ & $6.4$ & $8.9$ & $8.02$ & $11.0$\tabularnewline
\hline
\end{tabular}
\caption{Cross sections  for hadronic molecule
 candidates at LHC energies considering the theoretical decay rates predicted in Refs.  \cite{mol1,mol2,mol3}. }
\label{III}
\end{center}
\end{table}

Finally, in Table \ref{IV} we present our predictions for mesons which are glueball candidates, i.e.,  particles dominantly made of gluons. It is important to emphasize that none of them was up to now unambiguously identified.
However, the existence of glueballs is predicted in many theoretical calculations, including lattice QCD (For a recent review see, e.g., \cite{review_glueball}). In our calculations we use the two-photon decay rates proposed in Ref. \cite{magno_mario}, where the glueball production was estimated in ultraperipheral heavy ion collisions. 
Due to the small values of $\Gamma_{\gamma \gamma}$ for glueball states, we predict a low value for the total cross section and event rates smaller than $10^4$ events/year.

\begin{table}[t]
\begin{center}
\begin{tabular}{|c|c|c|c|c|c|}
\hline 
State & $\Gamma_{\gamma\gamma}^{th.}$(eV) & \multicolumn{2}{c|}{$\sigma_{Electric}$ (pb)} & \multicolumn{2}{c|}{$\sigma_{DZ}$ (pb)}\tabularnewline
\cline{3-6} 
 $X$ & $X\rightarrow\gamma\gamma$ & $7$ TeV & $14$ TeV & $7$ TeV & $14$ TeV\tabularnewline
\hline
\hline 
$f_{0}(1500)$ & $0.77$ & $0.080$ & $0.088$ &  $0.066$ & $0.10$\tabularnewline
\hline 
$f_{0}(1710)$ & $7.03$ & $0.38$ & $0.51$ & $0.46$ & $0.61$\tabularnewline
\hline 
$X(1835)$ & $0.021$ & $0.0009$ & $0.0012$ & $0.0011$ & $0.0014$\tabularnewline
\hline
\end{tabular}
\caption{Cross sections and event rates for glueball candidates at LHC energies considering the theoretical decay rates presented in Ref.  \cite{magno_mario}.  }
\label{IV}
\end{center}
\end{table}

Some comments are in order before the summary of our main results. Firstly,  
meson production in two-photon interactions are clean events, with a final state characterized by two very forward protons, the meson (or its decay products) in the central detector and the presence of two rapidity gaps.
Two rapidity gaps in the final state also are generated in central exclusive processes (CEP) by Pomeron - Pomeron interactions, where a Pomeron ($\pom$) is associated to a colorless object (See, e.g., \cite{forshaw}). 
The magnitude of meson production in $\pom \pom$ interactions has been estimated in Refs. \cite{antoni,magno,harland1,harland}. In particular, in Ref. \cite{harland1}, the central exclusive heavy quarkonia ($\chi$ and $\eta$) production at the LHC has been studied in detail. In comparison with our predictions for $\chi_{b,c}$ production, the results presented in \cite{harland1} are at least three orders of magnitude larger. It is important to emphasize that the CEP predictions describe the CDF data for the exclusive $\chi_{c0}$ production \cite{cdf}.  In the case of $\eta_{b,c}$ production, the CEP predictions are a factor two larger. Consequently, we expect that for these final states the production to be dominated by $\pom \pom$ interactions. On the other hand, the 
production of light mesons in $\pom \pom$ interactions still is an open question, since in this case the
main contribution for the cross section comes from non-perturbative regime, which implies the use of phenomenological models in order to estimate the total cross section  (For some related studies  see, e.g., Refs. \cite{antoni,magno,harland}). Moreover, $\pom \pom$ cross sections are strongly dependent on the treatment of the soft final states interactions and the associated survival probability.  In contrast, the predictions for light meson production in $\gamma \gamma$ interactions are much less sensitive to these effects. They are under theoretical control and can be considered a lower bound for the event rates of these final states. Another aspect which we would like to comment is the experimental separation among $\pom \pom$ and  $\gamma \gamma$ interactions.  For both cases  the $t$-distribution is of the type $\exp(-bt)$. The main distinction is associated to the  slope $b$ which is almost 4 (40) GeV$^{-2}$ for  
$\pom \pom$ ($\gamma \gamma$)  interactions \cite{forshaw,tagging,denterria}. It implies that photon - induced interactions  take place larger impact parameters (i.e. are less central) than Pomeron induced processes and, consequently,  the  exchanged squared  momentum  is smaller. 
As a consequence it is expected that the  transverse momentum distribution of the scattered protons to be different   $\gamma \gamma$ and $\pom \pom$ interactions, with the latter predicting  large $p_T$ values.  This expectation is corroborated by the results presented, for instance,  in the Refs. \cite{tagging, kepka}, where the transverse momentum distribution has been quantified considering different final states. 
  Certainly,  this subject deserves more detailed studies. However, it is important to emphasize that the   
ATLAS and CMS collaborations have a program of forward physics with extra detectors located in a region away from the interaction point, which probably can eliminate many serious backgrounds.

In summary,  in this letter we estimated the the meson production in two-photon interactions 
at CERN-LHC, which is characterized by two rapidity gaps in the final state. We have 
obtained non-negligible values for the total cross sections, which implies that 
the experimental study is feasible. In particular, our results indicate that this 
process can be useful to test the glueball and  hadronic molecule models.

\section*{Acknowledgments}
This work was supported by CNPq, CAPES and FAPERGS, Brazil.

\end{document}